# Exchange coupling and enhancement of Curie temperature of the intergranular amorphous region in nano-crystalline duplex-phase alloys system[1]

Y. Z. Shao[2], W. R. Zhong, G. M. Lin and X. D. Hu

*Department of Physics, Zhongshan University, Guangzhou 510275, China*

We explored the magnetic behavior of a common two-phase nanomagnetic system by Monte Carlo computer simulation of a modified Heisenberg model on a 3D complex lattice with single- and cluster-spins. The effect of exchange coupling between two component magnetic phases was studied on the enhancement in Curie temperature $T^A_c$ (ECT) of the intergranular amorphous region of a common duplex-phase alloy system, with numerous nano-crystallites embedded in amorphous matrix. The dependences of ECT were investigated systematically upon the nanocrystallite size $d$, the volume fraction $V_c$ and the interspace among crystallites $\xi$. It was observed that large crystallized volume fraction $V_c$, small grain size $d$ and thin inter-phase thickness $\xi$ lead to the obvious ECT of intergranular amorphous region whereas the Curie temperature of nanocrystallites $T^{cr}_c$ declines slightly. There is a simulative empirical formula as below: $T^A_c/T_c \sim 1/\xi$, which relates the reduced ECT to microstructure parameter $\xi$ and conforms to its experimental counterpart within an order of magnitude. In addition, we also simulated the demagnetization of a hard-soft nanocomposite system. We estimated the influence of exchange coupling between two component phases on the cooperativity of two-phase magnetizations and the coherent reversal of magnetizations as well as coercivity and energy product.

## I. Introduction

The magnetic system consisting of nanoscale duplex phases and epitaxial multilayer, if their microstructures of component phase are controlled properly, usually gains an advantage over single phase system in their magnetic behaviors and applications. The successful achievements are observed numerously in nanomagnetic systems of both soft magnetic materials like commercial Finemet [1~3] and hard magnetic materials such as Nd-Fe-B complex with α- Fe crystallites [4~7]. It is of considerable theoretical interest to study in detail the exchange coupling between the nanoscale grains (or layers), and between nanoscale grain and adjacent amorphous matrix as well as the effect of exchange coupling on the magnetic behavior of a nanocomposite. For a nanocomposite with two magnetic phases, it is also of practical interest to investigate systematically the influence of grain size $d$ (or thickness of layer), volume fraction of crystallite phase $V_c$ and interspace spanning between two coupled grains $\xi$ on microscale coupling interaction as well as macroscale magnetic characteristics, such as spontaneous magnetization, Curie temperature and coercivity etc. Yavari et al [2] and Makino et al [3] investigated the enhanced Curie temperature (ECT) within the intergranular amorphous region of duplex-phase Finemet alloy, and they ascribed the ECT to the inhomogeneous distribution and the diffusion of element Niobium and Boron. Different from the viewpoint of inhomogeneity above, Hernando et al [8~9] firstly dealt with the ECT of a duplex-phase alloy system on a base of pure magnetism instead. Hernando et al, after deducting the possible contribution of inhomogeneous distribution of element to ECT, proposed a phenomenological mechanism that the existence of a molecular field around 80T originated from the exchange interaction of α-Fe crystallites embedded within an amorphous matrix brings about the ECT mainly. Hernando et al also elaborated the contribution to the ECT from the inhomogeneous distribution and diffusion of element negligible if the interspace $\xi$ becomes smaller than 5 nm. The phenomenological formula given by Hernando et al relates the ECT with the interspace $\xi$ in the form of Eq.1.

---

[1] Supported by the Natural Science Foundation of Guangdong Province, P.R.China (No.021693)

[2] Correspondent author, email: stssyz@zsu.edu.cn

$$T_c^a = T_c^{a*} + (T_c^{cr} - T_c^{a*}) 2l / \xi \tag{1}$$

where $l$ and $T_c^{cr}$ symbolize a phenomenological exchange penetration constant about 5 angstrom and the Curie temperatures of crystalline α-Fe phase. $T_c^a$ and $T_c^{a*}$ stand for two kinds of Curie temperature of amorphous phases, i.e. that of intergranular amorphous region itself and that of an as-prepared amorphous ribbon with an identical composition like amorphous region. In their work, Hernando et al employed $T_c^a - T_c^{a*}$ as the indicative ECT of intergranular amorphous region. Hernando et al, nevertheless, have not specified any changes about the relevant Curie temperature of nanocrystalline α-Fe phase $T_c^{cr}$ of their investigating duplex-phase alloys in their original papers. The $T_c^{cr}$ we retrieved by Eq.1 using their experimental data, however, fluctuates remarkably for a small variation in composition and microstructure. A calculated $T_c^{cr}$ as high as 1094°C by Eq.1, moreover, is unreasonably large for a nanoscale α-Fe phase too. The formula (Eq.1) they proposed seems unable to be well self-consistent with their experimental data. Is the abnormality in $T_c^{cr}$ caused by the errors of their experimental data or simply the imperfection of phenomenological formula Eq. 1 itself? Skomoski also expressed somewhat his disbelief to those large values of appreciable ECT in his recent review on nanomagnetics [10].

Besides the reason mentioned above, two other reasons listed below motivate us to re-explore the ECT of a duplex-phase alloy system in this paper. Firstly, it is well-known that mean-field theory (molecular field) probably results in observable error nearby Curie temperature as well as the weakness of phenomenology itself. Secondly, the traditional top-down micromagnetic computation, based on classical magnetism model and conventional material parameters, probably causes considerable errors and becomes implausible when the grain size of materials is within a range of mesoscale, especially nanoscale. Aharoni has ever expounded some typical misinterpretations and mistakes in theoretical approach of some micromagnetics and one may refer to his review article [11].

Inspired by the idea of ab inito computation from single atom to clustered atoms, in this work we investigated the ETC and coercivity of duplex-phase alloy system once more by constructing in a bottom-up way a complex lattice which contains two species of spin sites, namely single-atom spin and nanoscale cluster spins, simultaneously. Moreover, without resorting to any phenomenological parameters of conventional bulk status, we performed the numerical computation by Monte Carlo simulation on a strict exchange coupling of Heisenberg model rather than a mean-field approach employed by previously reported work. We are curious about what a difference, if any, to be possibly made by the two distinct approaches. Note that only reduced parameters without any units were adopted in present simulation, the results obtained here are probably applicable to a variety of duplex-magnetic phase alloys instead of a specific one. To our knowledge, few researches have brought into such a comparison between the two distinct methods on this topic. This paper was organized in five sections. We gave this brief introduction in section 1 and described the relevant algorithm of Monte Carlo simulation and the experimental procedure of measuring magnetic property of Finemet in section 2. We presented in section 3 both simulative and experimental results as well as a relevant discussion in section 4. Finally, a summary was given in section 5.

## II. Descriptions of Model and Computational Algorithm
### 2.1 Modified Heisenberg model and other simulation parameters

In order to simulate the behavior of a duplex-magnetic phase spin system, we modified the Hamiltonian of a classical Heisenberg spin system (the first term in Eq.2a) by introducing both determinative uniaxial single-ion anisotropy parallel to the z-axis (the second term in Eq.2a) and random uniaxial anisotropy (the third term in Eq.2a ) in the form as shown in Eq.2a. The fourth term is Zeeman energy with an external driving field parallel to Z axis.

$$\hat{H} = -(J_{NN} + J_{NNN}) \sum_{<i,j>} (\vec{S}_i \cdot \vec{S}_j) - A \sum_{i'} (\vec{S}_{i'}^z)^2 - D \sum_{i''} ((\vec{S}_{i''} \cdot n_{i''})^2 - 1) + g\mu H_Z \sum_i \vec{S}_i^Z \quad (2a)$$

where $\quad \vec{S}_i \cdot \vec{S}_j = c_x S_i^x S_j^x + c_y S_i^y S_j^y + c_z S_i^z S_j^z \quad (2b)$

Symbol $S_i$ and $S_j$ in Eq.2 represent the spin at site $i$ and neighboring site $j$, respectively, within the lattice. $S^x$, $S^y$ and $S^z$ in Eq.2b denote, in turn, the projections of spin $S$ along x-, y- and z-axes. The spin-exchange constants $J_{NN}$ and $J_{NNN}$ signify the interaction of site $i$ with its nearest neighbors (NN) and the next-nearest neighbors (NNN), respectively. Depending on whether $J_{NN}$ or $J_{NNN}$ is selected, the summation $\Sigma_{<i,j>}$ symbolizes the sum over NN or NNN site pairs, respectively. Parameters $c_x$, $c_y$ and $c_z$, ranging from 0 to 1, are anisotropy constants of spin exchange. The isotropic Heisenberg spin system corresponds to the case when $c_x = c_y = c_z = 1$ and this is the case we handle in current project. On the other hand, if $c_x = c_y = 0$ and $c_z = 1$ the Hamiltonian describes the anisotropic Ising spin system. Four types of energy terms in the components of the Hamiltonian in Eq.2a account for the spin exchange energy of coupling between single-spins or cluster-spins, determinative uniaxial single-ion anisotropy energy, random uniaxial anisotropy energy and interaction energy between spin and external field, respectively. The spins in the first term include all possible sites throughout the entire 3D lattice and the first term governs the spontaneous magnetism of system. The counterparts, however, of the second and the third terms only occupy those sites belonging to either cluster spin sites (site $i'$ ) of crystallite part with a determinative orientation to a easy axis (Z axis) or single-atom spin sites (site $i''$ ) of amorphous matrix part with numerous random orientations, respectively. The $n_{i''}$ denotes a unit vector independently chosen for each site $i''$ with a random local easy direction which varies from site to site. Assuming the percentages of cluster-spin site $i'$ and single-spin site $i''$ to be $x$ and $1-x$ respectively, we can construct a complex lattice site with a continuous variation of x value. Depending upon the condition of simulation, it is available for us to adjust the size $d$ of cluster-spin site $i'$ in computation too. Considering the configuration of spin alignment, three kinds of spin exchange constants were defined for this complex lattice respectively, i.e. $J_{aa}$, $J_{bb}$ and $J_{ab}$, signifying the direct coupling intensities and cooperative capability among those spins within amorphous and crystalline phases ($J_{aa}$, $J_{bb}$) as well as the exchange coupling intensity of interface spins between amorphous matrix and crystalline phase ($J_{ab}$). According to the local spin-site and spin–pairs in 3D lattice scanned by Monte Carlo simulation, the program can identify the microenvironment of lattice and decide which one of three exchange constants to be selected. Considering the situation of a duplex soft-soft magnetic phase alloy system, we limited the anisotropy constant $A$ and $D$ in Eq.2a to small values, i.e. with both a small crystallization anisotropy of a crystallite and small random anisotropy of amorphous matrix. Through the combinations of some other larger parameters $A$ and $D$ of the Hamiltonian of Eq.2, we could also readjust the anisotropies of lattice and simulate the magnetic behaviors of other duplex-magnetic phase alloy system, such as duplex hard-hard magnetic system or hard-soft magnetic one. In light of the theory of superparamagnetism that all single-atom spins within a cluster align in identical direction, we could work out the spin of a cluster-spin as $S_c = d^3 S_a$, where $S_a$ is the spin of single-atom. The 3D lattice (size N with $N^3$ sites) of simulation is comprised of numerous small basic cubic sub-lattices and cluster-spin size $d_c$ is measured in a unit of a basic cubic cell consisting of four single atoms. For the convenience of comparison between the simulation result and experimental data, we also presume that a basic cubic cell owns a dimension equivalent to 2.5 angstrom, quite close to the real value of Finemet 2.8 angstrom [1].

The most important reduced parameter $k_B T_C/J_{NN}$, the ratio of the critical temperature against the exchange interaction, was calculated using Eq.3 [12],

$$k_B T_C / J_{NN} = 5(R - 1)[11S(S + 1) - 1]/96 \tag{3}$$

where $R$, $S$, $k_B$ and $T_c$ are the number of the nearest neighbors, lattice spin, Boltzmann constant and critical temperature, respectively. The interaction of site $i$ with the next-nearest neighbor $J_{NNN}$, which drops exponentially with respect to the distance between lattice sites[13], can be determined from $J_{NN}$ and is usually taken as 0.1~0.25 $J_{NN}$ [14]. In current paper, we take $J_{NNN}$ to be 0.2 $J_{NN}$. The critical temperature $T_c$ was set to 575 °C in reference to the Curie temperature of Finemet [1].

**2.2 Algorithm of Monte Carlo simulation**

In Monte Carlo (MC) simulations the orientations of the lattice spins can be initialized either orderly, such that all lattice spins align parallel to z-axis, or randomly. We employed random initialization to avoid the influence of retained order on the final results. Periodic boundary condition was used to minimize the influence of the finite lattice size. Spherical coordinate system was adopted with symbols $\psi$ and $\theta$ signifying the azimuthal angle of the final average moment and the angle the final average moment made with the z-axis respectively. A lattice site $i$ with spin orientation $S_i(\theta_i, \psi_i)$ was chosen randomly and the value of the Hamiltonian was calculated with Eq.2 as $H_i$. The spin on this lattice site was then rotated by thermal activation to a new random orientation, $S_n(\theta_n, \psi_n)$ and the new value of the Hamiltonian $H_n$ of site $i$ was similarly calculated with Eq.2. The energy variation caused by such a change of spin orientation is

$$\Delta H = H_n - H_i \tag{4a}$$

The Metropolis criterion [15] was adopted to decide whether the new spin orientation is accepted or rejected. The energy change must satisfy the condition below for a successful spin orientation change:

$$\exp[-\Delta H/k_B T] \geq Q \in (0,1) \qquad (4b)$$

where $Q$ is a random number within the range from zero to one.

The simulation proceeded by sweeping every lattice site in sequence for a number of repetitions (Monte Carlo steps, MCS) and the statistical averages of the magnetic properties concerned were computed over ten independent simulations. Standard tests were performed to verify whether equilibrium under the prescribed condition was attained. All simulations were performed on a three-dimension lattice with a periodic boundary condition and lattice size N=60~120. In our simulation the sweeping times counted up to $10^5$ MCS.

**2.3 Magnetization and other micro-structural parameters concerned in this paper**

In our simulation, we kept the spin system in contact with an isothermal heat bath at temperature $t$. With $\vec{S}_{i'}$ and $\vec{S}_{i''}$ denoting the magnetic spins occupying site $i'$ and site $i''$ respectively, the magnetization $m$ averaged over all lattice sites is given by

$$\vec{m} = \frac{1}{N^3} \sum_{i \in i'+i''} (\vec{S}_{i'} + \vec{S}_{i''}) \qquad (5)$$

The temperature dependences of magnetization, $m \sim t$, with various volume fractions of crystallization $x$ and crystallite sizes $d_c$ were simulated systematically and the Curie temperature was ascertained at $m = 0$. Two types of Curie temperature, i.e. $T_c^{cr}$ for crystallization phase and $T_c^A$ for amorphous matrix, were defined for the duplex-magnetic phase alloy system. Unlike that adopted by Hernando et al, namely $T_c^a$ - $T_c^{a*}$, we employed a reduced form $T_c^A/T_c^{cr}$ to quantify the ECT of duplex-magnetic phase alloy system, and this is more convenient and accurate in simulation than introducing a reference amorphous phase with a same composition as investigated intergranular amorphous matrix region. We defined in this paper the interspace $\xi$ similar to that of Hernando et al [8, 9] to ensure a comparison of equality with the result early reported. Eq.6 was used to calculate the parameter $\xi$ under the circumstances of both experiment and simulation. In the case of experiment, we simply substituted the experimentally measured volume fraction $V_c$ of α-Fe crystallite for the counterpart $x$ set in simulation.

$$\xi = d_c(1/x^{1/3}) - d_c \qquad (6)$$

In addition to the spontaneous magnetization at zero-field, the demagnetization curve of a hard-soft nanomagnetic system was simulated tentatively at nonzero-field and the different intrinsic coercive features were revealed.

**2.4 Experimental procedure**

In order to testify the computer simulation, we also carried out relevant measurement of magnetic property of duplex-phase alloys with various the volume fractions and grain sizes of crystallites. Commercial Finemet amorphous ribbons ($Fe_{73.5}B_9Nb_3Cu_1Si_{13.5}$) were annealed in vacuum state at different temperatures and times (partial crystallization process) to gain various stages of nanocrystallization, and duplex-phase microstructures with various volume fractions and grain sizes of crystalline phase were prepared for measurement. The annealing process is the temperature range between 460°C~620°C with the time of 30min~60min. The microstructure of nanocrystalline state was characterized by means of conventional x-ray diffraction carefully. The grain size $d$ was estimated by means of the Scherr formula from the width of the x-ray diffraction pattern after subtracting the contribution of instrumental broadening. The volume fraction of nanometer α-Fe crystallites was determined approximately from the relative x-ray intensities of crystalline grain $I_c$ and amorphous matrix $I_a$ using a formula below.

$$V_C = I_c / (I_c + K \cdot I_a) \tag{7}$$

where $K$ is a constant about 0.9 ~1. The measurement of x-ray diffraction was conducted using a Rigaku D/Max-IIIA diffractometer. The temperature dependence of magnetization and Curie temperature of duplex-phase Finemet were measured using a magnetic thermogravimeter (Netzsch TG-209).

**III. Results**

Figure 1 shows some typical x-ray diffraction patterns of Finemet in as-quenched state and annealed at various temperatures after 60 minutes. α-Fe crystallites grew out of original amorphous matrix and at an annealing temperature $T_a$ =540°C most parts of amorphous matrix have transformed into crystalline grains with a average size about 20~25nm. Figure 2 displays the average grain size $d$ and the volume fraction $V_c$ of α-Fe crystallites after different annealing processes, which prepared an ideal duplex-phase microstructure with a diverse combination of $d$ and $V_c$ for magnetic measurement. Figure 3a exhibits some typical magnetic thermogravimetric (MTG) results of Finemet in both as-quenched and different annealed states. The typical two-stage trend with an inflexion on the general MTG curve of nanoscale α-Fe crystallite and amorphous matrix was observed, characteristic for a duplex-magnetic phase system. The relevant Curie temperatures of both crystalline and amorphous phase were detected by magnetic thermogravimeter from the measured differential MTG curves. We marked the location of relevant Curie temperature of crystalline and amorphous phase with two vertical-arrow indicators in figure 3. The Curie temperatures of as-quenched material measured by us are 312.6 °C and 582.2°C for amorphous and crystalline states, respectively. The two Curie temperatures above are close to those of amorphous Fe-Si-B-M and nanocrystalline Finemet early reported by Yoshizawa et al. [1]. It is worthy to notice the changing curvature at the inflexion of MTG curves near $T_c^A$. The clear-cut inflexion at initial crystallization stage becomes gradually indistinct with crystallization process and smeared finally. And this trend of magnetization versus temperature is regarded as the transition from a inhomogeneous two-phase mixture into a two-phase nanostructure [10], indicating that α-Fe crystallites are more correlated magnetically with residual amorphous matrix on a nanoscale and a strong exchange coupling exists between two magnetic phases. Interestingly, an analogical trend in the variation of reduced magnetization $M/M_s$ versus reduced temperature

$T/T_c$ was also observed clearly in our Monte Carlo simulation, as displayed in figure 3b. Obviously, the crystallization of α-Fe crystallite gives rise to an enhancing Curie temperature of remnant amorphous matrix region $T^A_c$. Further systematic simulations under the conditions of various cluster sizes $d$ and volume fractions $x$ (or $V_c$) of the duplex-magnetic phase system were carried out and figure 4 demonstrates the reduced Curie temperature $T^A_c / T_c$ versus both $d$ and $x$. To verify our simulation, we also made in figure 4 a direct comparison between the present simulation result and our experimental data points (square symbol) as well as preceding ones (circle symbol) reported by Hernando et al [8, 9]. For the sake of convenience to guide the eyes of readers, the graphs at two different viewpoints were provided in figure 4. Most of our experimental data points could find their positions on or scatter in the vicinity of the simulative data surface while parameter $d$ and $x$ vary within a wide range. Those experimental data points we retrieved from the work of Hernando et al [8, 9], due to the original fluctuation and quite larger value in $T^{cr}_c$, turn out to be a smaller and discrepant reduced ECT ($T^A_c/T^{cr}_c$) while compared with our simulated data surface. Figure 4 indicates visually the enhancement of $T^A_c$ and its dependence upon the grain size $d$ and the volume fraction $x$ of α-Fe crystallites. It is necessary to make clear here that we adopted in figure 4 one nanometer as the unit of grain size $d$ instead of the basic cubic lattice cell unit in simulation. According to Eq.6, we figured out the relevant average interspace $\xi$ of remnant amorphous matrix among α-Fe crystallites under the various conditions of grain sizes $d$ and volume fractions $V_c$. Table 1 lists our experimental data concerning Curie temperature and other microstructure parameters. For the purpose of comparison, we also cited some original data of the duplex-phase alloy Fe~B~Nb~Cu reported by Hernando et.al [8, 9] in Table 2. It is necessary to point out that the Curie temperature of nanoscale α-Fe crystallite $T^{cr}_c$ and $T^a_c / T^{cr}_c$ added in table 2 were retrieved by us using their original data and proposed formula Eq.1 latter. Figure 5 presents the variations of both $T^A_c$ and $T^{cr}_c$ versus the average intergranular space $\xi$. With the decrease of $\xi$, the Curie temperature of amorphous matrix $T^A_c$ raises whereas its counterpart of nanoscale α-Fe crystallite declines slightly. Evidently in Finemet, the extent of ECT (30~60°C) in current paper is smaller than that of Hernando et al (50~110°C), and the trend of $T^{cr}_c$ is also dissimilar from that of Hernando et al in that the $T^{cr}_c$ in our case shows a regular decrease with shortening $\xi$. We plotted $T^A_c/T^{cr}_c$ against the reciprocal $\xi$ out of simulative and experimental results in figure 6, respectively. Apparently, reduced ECT $T^A_c / T^{cr}_c$ maintains a good linear relationship with the inverse $\xi$, and this is quite similar to the Eq.1 proposed by Hernando et al. in spite of the difference in the definition concerning the ECT. In our work, we adopt reduced ECT $T^A_c / T^{cr}_c$ in stead of $T^a_c - T^{a*}_c$ by Hernando et al.

  Finally, in order to reveal the demagnetizing characteristics, more precisely the degree of coherent reversal and cooperative magnetization, of a common hard-soft nanocomposite and the influence of exchange coupling between two component phases on the magnetic behavior, we also simulated the demagnetization curves and differential susceptibilities of a common hard-soft duplex nanomagnetic system under the different conditions of exchange coupling constants and cluster sizes. Contrast to the situation of preceding soft-soft duplex system in which only single-species spin is involved, two sorts of different spins are included for the hard-soft duplex system. Figure 7a shows one of some typical results with a sharp two-phase-like inflexion on the demagnetization curve observed, and a less coercivity and substantially smaller magnetic energy product are obtained when exchange coupling constants reduce from $K_BT_C/J = 0.5$ to 1.8. The maximal differential susceptibilities, however, locate at an identical external field, indicating the occurrence of the magnetization reversal of soft-phase without the influence from variation of $J$ value. The two-phase-like inflexion diminishes with decreasing cluster size (or interspace $\xi$) and disappears completely while $d$ =1, as shown in figure 7b. No a significant variation in coercivity and energy product was observed in our simulation for the smaller ($d=1$) and larger ($d=8$) cluster ensembles, and the differential susceptibility peak of soft-phase vanishes absolutely in the case of fully exchange coupling, dissimilar from the situation in figure 7a. In addition, as displayed in figure 7, the system retains steadily the state of a broad switch-field distribution owing to both isotropic Heisenberg spin system adopted and such great difference in anisotropy as $A/D$ =1000 introduced for relevant hard and soft phase.

**IV. Discussion**

Alben et al [16] tackled the magnetic behavior of single-phase amorphous alloys in terms of random anisotropy model, which has been serving as the base of analyzing the magnetism of amorphous alloys. Based upon the random anisotropy theory and through a simple statistical mechanics approach, Herzer studied the dependence of anisotropy energy <$E$> (or coercivity $H_c$) upon the grain size $d$ after the complete crystallization of a single amorphous alloy system [17], and he put forward a concise power-law relationship concerning the anisotropy energy <$E$> with grain size $d$, viz. <$E$> ~ $d^6$. In contrast to the approximate statistical approach, the theoretical computations by Fisch [18] on Monte Carlo simulation of Heisenberg model with random anisotropy have given great inspiration to our work, and we based on it the current simulation of the dilute Heisenberg model with two species of spins. Actually, one of our early studies [19] on transition from superparamagnetism to ferromagnetism in a Heisenberg model regarding a single-species cluster-lattice has found that anisotropy energy <$E$> (or coercivity $H_c$) and cluster size $d$ obey a power law with a changing exponent $P$, i.e. <$E$> ~ $d^P$. The exponent $P$ is not a universal constant equal to 6 as Herzer has ever suggested, but a variable depending upon system temperature $t$ and uniaxial anisotropy constant $A$ in Eq.2. We further devoted our efforts to a duplex-phase system with either single species or two different species spins in this work.

For a duplex-magnetic phase system consisting of a high-Curie temperature phase (HCTP, e.g. the nanoscale α-Fe crystallite) and a low-Curie temperature phase (LCTP, e.g. amorphous matrix), respectively, the mechanism underlying the enhancing Curie temperature of LCTP is in principle ascribed to either the indirect ferromagnetic interaction of two HCTP crystallites via LCTP region or the direct exchange coupling between HCTP crystallite and adjacent LCTP matrix on their interface. Both the ferromagnetic interaction and exchange coupling mentioned above introduce an extra magnetic ordering within the LCTP region and this gives rise to the enhancement of Curie temperature of LCTP matrix, no matter LCTP is either soft magnetic [8, 9] or hard magnetic [20]. The shortening of intergranular space of amorphous matrix $\xi$ intensifies the ECT of amorphous phase as it was ever expounded in Hernando et al work [8, 9]. One could refer to their relevant work and we would not repeat their explanation further here. In this section we focus on the new features we have obtained in experiment and simulation, namely smaller ECT than that of Hernando et al. and the slight decline of $T^{cr}_c$ with shrinking $\xi$ observed in our experiment as well as the demagnetization characteristics of hard-soft nanomagnetic system in simulation.

The figure 3a displays clearly the existence of an inhomogeneous two-phase mixture at early crystallization stage. Does the inhomogeneity contribute much to the ECT of amorphous matrix as Yavari et al. [2] and Makino et al. [3] suggested? The approximate consistency of our measured $T^a_c/T^{cr}_c$ with that estimated by our simulation (without involving in any inhomogeneity and diffusion) indicates that the inhomogeneity in duplex-phase Finemet probably plays a minor role in the ECT of amorphous matrix. And experimental results of the smaller intergranular space $\xi$ (< 5nm) and ECT extent (30~60°C) in table 1 also support the above judgment concerning the effect of inhomogeneity on ECT. Hernando et al has ever identified respectively the contributions of both element factor and pure magnetic interaction to the ETC in their work of duplex-phase alloy FeBNbCu [8, 9]. They, after deducting Cuire temperature increment of 20~50°C contributed by element factor from their measured one about 50~110°C, ascribed the ETC around 30~50°C to the contribution of pure magnetic exchange coupling when intergranular space $\xi$ is less than 5 nanometer. Considering the tendency and extent of ECT, we also deem that the smaller intergranular space $\xi$ should be one of reasons responsible for the absence of the contribution of inhomogeneous distribution of elements and diffusion to the ECT in our currently investigated Finemet. In addition, smaller ECT value observed in Finemet probably has something to do with the difference on our definition of ECT, i.e. our $T^a_c/T^{cr}_c$ and their $T^a_c - T^{a*}_c$.

Another interesting finding of current investigation is that the experimentally measured Curie temperature of nanoscale α-Fe crystallite $T^{cr}_c$ decreases slightly with shortening intergranular space $\xi$ or increasing grain size $d$, which is quite different with what the work of Hernando et al implies. No drastic fluctuating and unreasonable high Curie temperature were observed in our experiment. Gradually declining $T^{cr}_c$ with $\xi$ probably hints the side-effect of reverse magnetic interaction of amorphous phase on neighboring crystalline phase and this seems to be neglected by other previous work. Unfortunately, our current simulation computation failed to reveal the above subtle influence of amorphous matrix on $T^{cr}_c$ owing to the limit of our simulation software itself. $T^{cr}_c$ in our simulation program, once set a certain value at first, was then handled as a constant, and $T^a_c$ of amorphous phase was reduced against $T^{cr}_c$. It is well established in a single-phase nanomagnetic material that the decrease of grain size usually results in declining Curie temperature apparently [21]. As for a duplex-phase nanomagnetic system consisting of a LCTP and HCTP, however, the situation becomes complicated due to the mutual effect of two phases. In the case of strong exchange coupling, for instance, the duplex magnetic-phase system exhibits only one instead of two Curie temperatures as usually [22,23]. Furthermore, we think that presence of LCTP, whose Curie temperature is enhanced by adjacent HCTP, diminishes the Curie temperature of adjacent HCTP simultaneously. Taking into account the exchange coupling on the interface between amorphous matrix and α-Fe crystallite, asymmetry of atom-sites on the two sides of the interface makes it impossible for those α-Fe spins on the interface to couple in that way as their interior counterparts do, and this leads to the decrease of $T^{cr}_c$. The greater the percentage of the interface is, the more evidently the $T^{cr}_c$ declines. The shrink of interspace $\xi$ of intergranular amorphous matrix in a certain stage of crystallization possibly increases the interface. Besides the asymmetry of exchange coupling on the interface, the indirect ferromagnetic interaction of two α-Fe crystallites via amorphous matrix also affects the $T^{cr}_c$. For a uniaxial anisotropic granular nanomagnetic system, even though the easy axis takes a random orientation each other among these crystallites nearby, the collaboration of interactive magnetic moments could prevents the $T^{cr}_c$ from decline further if the interspace $\xi$ becomes quite small. Thus, we predict that the $T^{cr}_c$ declines only in a certain stage of crystallization and will rise when crystalline α-Fe phase dominates over the duplex-phase system in volume fraction and grain size. Further investigation on this aspect is necessary.

As we specified in section 2.1, there are three kinds of spin exchange constants $J_{aa}$, $J_{bb}$ and $J_{ab}$ in our simulation, and the exchange coupling between two phases depends upon the magnitude of $J_{ab}$. Changing the value of $J_{ab}$ and cluster size $d$, we can gain an insight into the influence of exchange coupling on the demagnetization of a hard-soft nanomagnetic system, as shown in figure 7. Generally, greater $J$ value, namely stronger spin exchange coupling, makes it more easily for the magnetizations of two magnetic phases to rotate coherently when a reverse external field is applied, and this leads to the fading-out and even complete vanishing of two-phase-like inflexion or shoulder on demagnetization curve. The similar mechanism also holds for the situation of smaller grain or cluster size $d$ if the size is less than exchange length $l_{ex}$ ($\sim \sqrt{A_J}/M_S$, where $A_J$ and $M_S$ are exchange stiffness and spontaneous magnetization, respectively). Theoretically, $l_{ex}$ ranges between 1 and 2 nm for most materials [10]. Due to the exchange coupling, nanoscale soft-magnetic phase is magnetically hardened by neighboring hard-magnetic phase, and the coercivity of a nanocomposite system has contributions from both hard and soft magnetic phases. The great difference in anisotropy of two magnetic phases (A>>D), however, scarcely enables the magnetizations of relevant two phases to reverse synchronously in a cooperative way, resulting in a broad switch-field distribution [24] as well as the appearance of characteristic two-phase-like shoulder on demagnetization curves. Does a broad switch-field distribution mean consequentially a non-cooperative magnetic reversal as the reference [24] claimed? Our simulative results in figure 7 show that the enhancement of exchange coupling between two nanomagnetic phases, either by increasing exchange coupling constant or decreasing crystallite size, could bring in the transition from an incoherent rotation of magnetization to a coherent mode, but the original state of broad switch-field distribution nearly remains intact. Actually, the shape of hysteresis loop or the characteristics of switch-field distribution are affected by three main factors in our simulation, namely the anisotropy of spin exchange, the combination of crystalline and random anisotropy, the type and direction of external field. And these factors are controllable through adjusting such relevant parameters as $J$, $A$, $D$ and $H$ in Eq.2a and $c_x$, $c_y$ and $c_z$ in Eq.2b. In regard to a two-phase nanomagnetic system, it seems more reasonable to take the disappearance of shoulder on demagnetization curve rather than switch-field distribution as the indicator of intergranular cooperative phenomenon.

**V. Summary**

We studied the magnetic characteristics of a duplex-phase nanomagnetic system by Monte Carlo simulation of a modified classic Heisenberg model involving in both determinate and random uniaxial anisotropies on a complex lattice. The lattice consists of both single and cluster spins built in a bottom-up way. Under the various conditions of volume fraction and cluster size of component spins, the temperature dependence of spontaneous magnetization was computed and two sorts of Curie temperatures were identified on the general curves of magnetization versus temperature. The experimental measurement was also carried out in a duplex-phase magnetic alloy to account for those simulation results. In addition to spontaneous magnetization and Curie temperature, we also simulated tentatively the demagnetization of a hard-soft nanocomposite system. We estimated the influence of exchange coupling between two component phases on the cooperativity of two-phase magnetizations and the coherent reversal of magnetizations as well as coercivity and energy product. Following are two main conclusions we drew from this paper.

1. The magnetic interplays in nanoscale, like exchange coupling and interaction between high-Curie-temperature (HCT) and low-Curie-temperature (LCT) phase, bring about a considerable enhancement of the Curie temperature of LCT phase and a sight decline of the Curie temperature of HCT phase simultaneously. An empirical formula was obtained in both experimental measurement and Monte Carlo simulation, which shows that the enhancement of reduced Curie temperature $T^A_c/T_c$ of remanent amorphous matrix maintains a good linear relationship with the inverse interspace $\xi$, $T^A_c/T_c \sim 1/\xi$. In contrast to magnetic interaction, the inhomogeneity and diffusion of elements simply play a minor role in their influence on Curie temperatures of component phases.

2. Both increasing exchange coupling between two component magnetic phases and decreasing size or interspace of nanoscale crystallites or clusters may yield a transition from incoherent magnetization reversal with two-phase-like characteristic to a cooperatively coherent mode. The decreases of grain size and interspace have a slight effect on coercivity and magnetic energy product whereas the increase of exchange coupling affects them strongly. The increase of exchange coupling, however, does not impact on the switch-field of soft-phase at all.

**References**

[1] Y. Yoshizawa, S. Oguma and K .Yamauchi, *J.Appl.phys.*, 64(1988) 6044

[2] A. R. Yavari and O. Drbohlav, *Mater. Trans. JIM.*, 26(1995) 896

[3] A. Makino, T. Hatanai, A. Inoue and T. Masumoto, *Mater. Sci. Eng.*, A226-228 (1997) 594

[4] T. Schrefl, R. Fischer, J. Fidler and H. Kronmuller, *J.Appl.Phys.*, 76(1994)7053.

[5] R. Skomski, *J.Appl.Phys.*, 76(1994)7059.

[6] Z. H. Cheng, H. Kronmuller and B. G. Shen, *Appl. Phys. Lett.*, 73(1998)1586.

[7] H.W. Zhang, S. Y. Zhang and B. G. Shen, *Phys. Rev. B.*, 62(2000) 8642.

[8] A. Hernando, I. Navarro and P. Gorria, *Phys.Rev. B.*, 51(1995)3281.

[9] A. Hernando and T. Kulik, *Phys. Rev. B.*, 49(1994)7064.

[10] R. Skomski, *J.Phys:Condens Matter*, 15(2003)R841

[11] A. Aharoni, *Physica B*, 306(2001)1.

[12] G. S. Rushbrooke and P. J. Wood, *Mol. Phys.*, 1(1958)257.

[13] D. Wagner, *Introduction to the theory of magnetism*, Pergamon Press, New York, 1972. p153.

[14] J. Mlodzki, F. R. Wuensch and R. R. Galazka, *J. Mag. Mag. Mater*, 86(1990)269.

[15] K. Binder and D. W. Heermann, *Monte Carlo simulation in statistical physics*, Springer, Berlin, 1992. p14.

[16] R. Alben, J. J. Becker and M. C. Chi, *J. Appl. Phys.*, 49(1978)1653.

[17] G. Herzer, *IEEE Trans. Magn.*, 26(1990)1397.

[18] R. Fisch, *Phys. Rev. B.*,58(1998)5684.

[19] C. H. Shek, Y. Z. Shao and J. K. L. Lai, *Physica A*, 276(2000)201.

[20] G. C. Hadjipanayis, *J. Mag. Mag. Mater*. 200(1999)373.

[21] C. Suryanarayana, *Inter. Mater. Rev.*, 40(1995) 41

[22] R. Skomski and D. J. Sellmyer, *J. Appl. Phys.*, 87(2000)4756.

[23]U. Bovensiepen, F. Wilhelm, P. Srivastava, P. Poulopoulos et al., *Phys. Rev. Lett.*, 81(1998)2368.

[24] R. Skomski and D. J. Sellmyer, *J. Appl. Phys.*, 89(2001)7263.

Figure 1. X-ray diffraction pattern of as-prepared and annealed samples.

Figure 2. The grain size and volume fraction of crystallites versus annealing temperature.

Figure 3a. Measured magnetic thermogravimetric (MTG) curves and differential MTG curves of duplex-phase Finmet in different crystallization stages.

Figure 3b. Simulated temperature dependence of spontaneous magnetization of the two-species spin system in different percentage of cluster-spins.

Figure 4. Reduced Curie temperature of amorphous phase under the various grain size and volume fraction of crystallites. Data surface was simulated by means of Monte Carlo method in current project, and the experimental data points by us (square) and Hernando et al (circle) were included for the purposes of comparison.

Figure 5. Measured variation of Curie temperatures of remnant amorphous and crystalline phase with the average interspace of amorphous region.

Figure 6. The reduce Curie temperature of amorphous phase versus inverse interspace by experiment (a) and simulation (b).

Figure 7. The demagnetization curves and differential susceptibility of a hard-soft nanomagnetic system simulated by means of Monte Carlo method. The inset is the relevant simulation conditions for the variations of exchange coupling constant $J$ (a) and cluster-spin size $d$ (b), respectively.

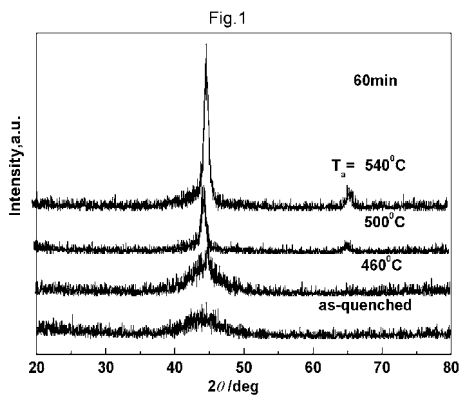
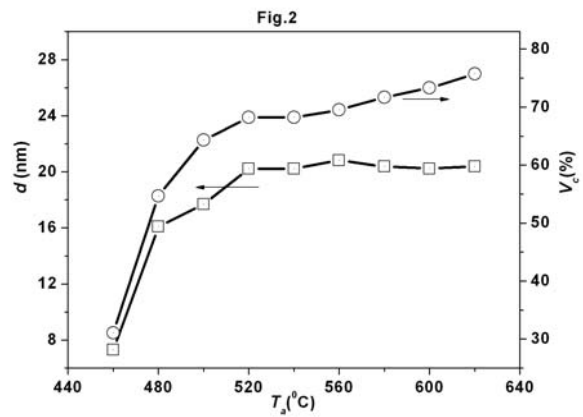

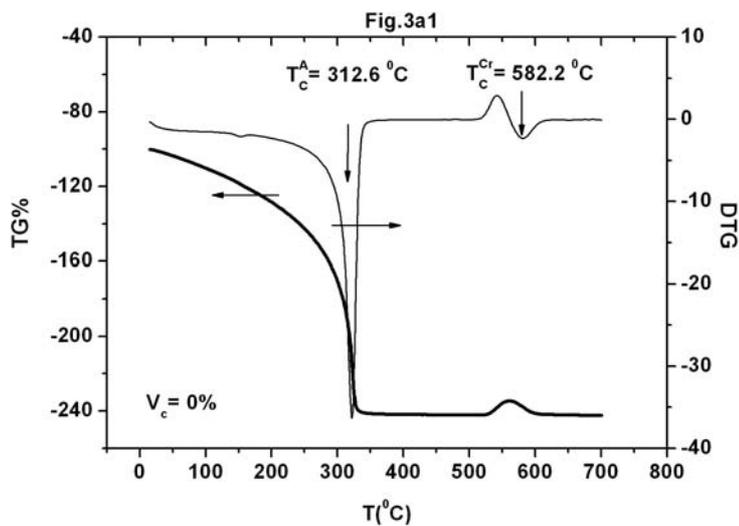

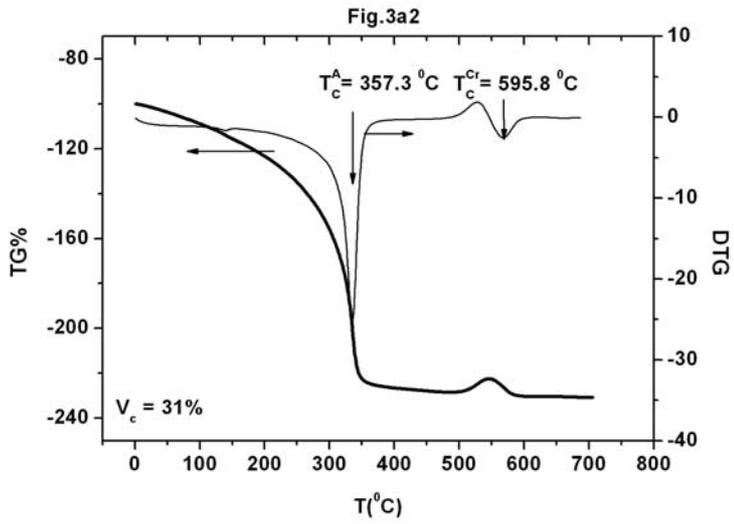

Fig.3a2

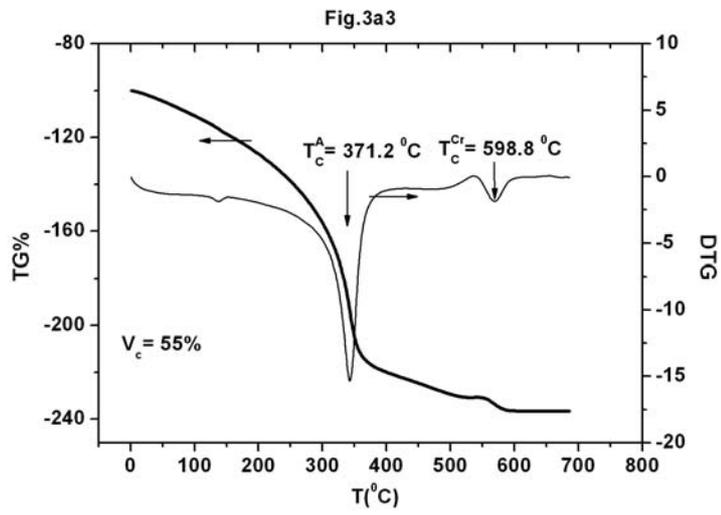

Fig.3a3

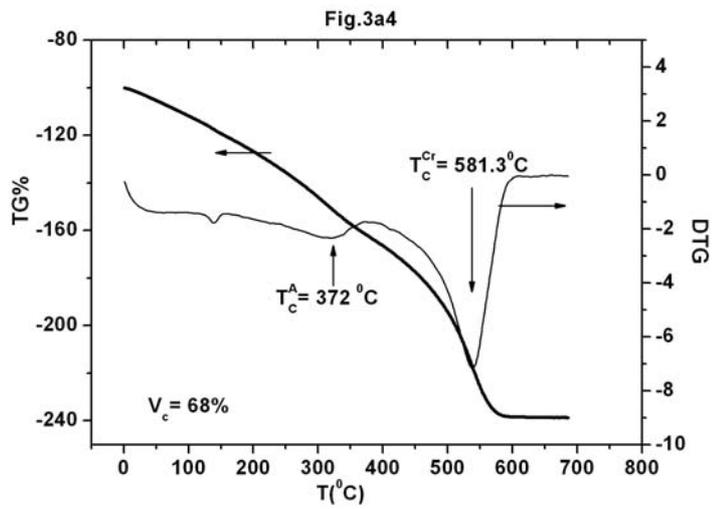

Fig.3a4

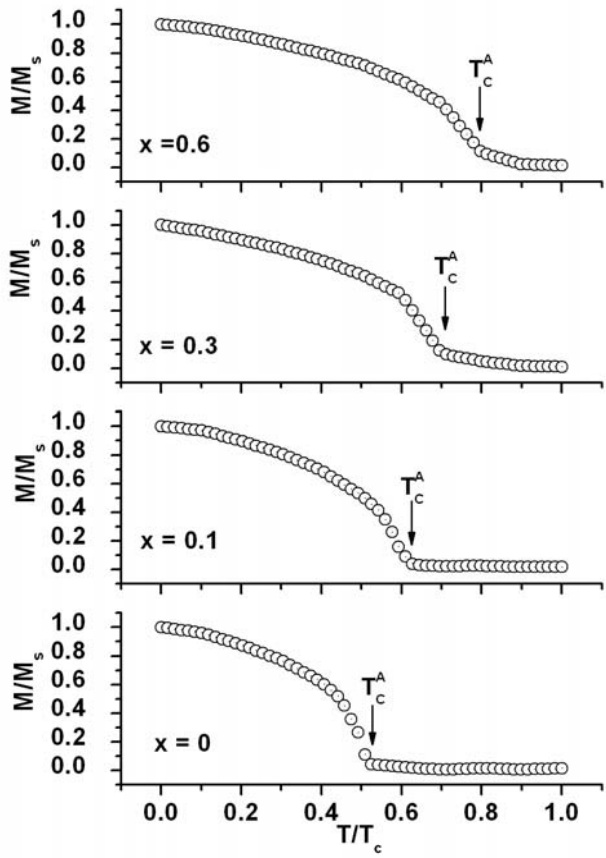

Fig.3b

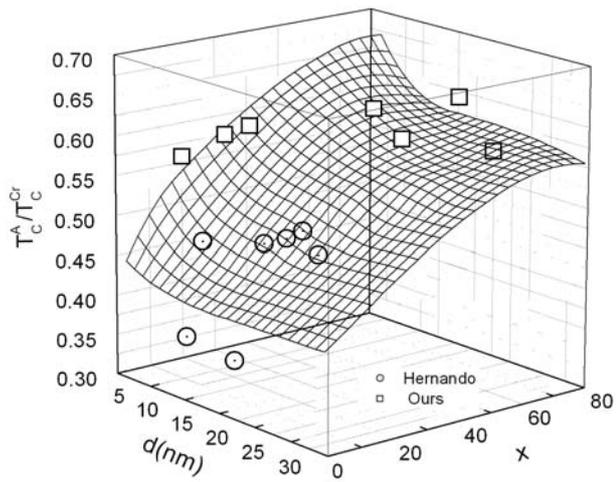

Fig.4a

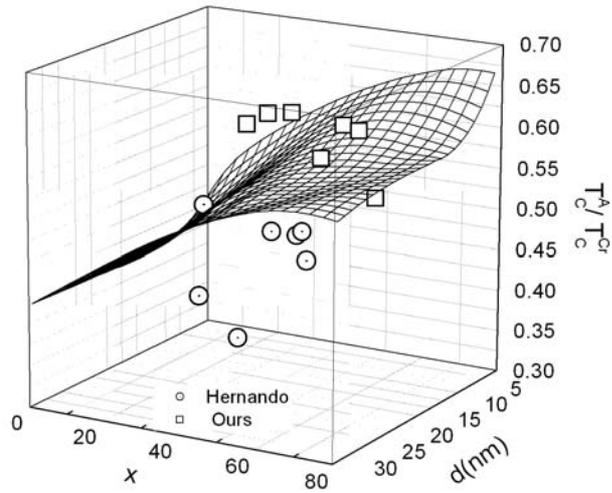

Fig.4b

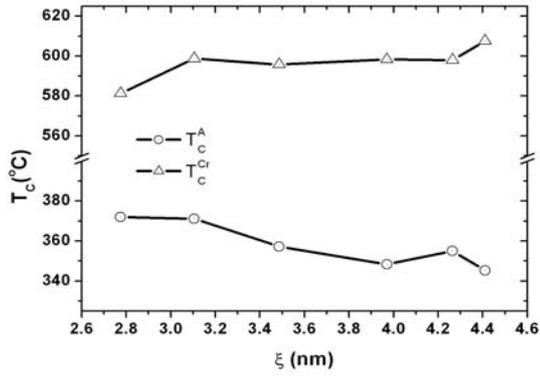

Fig.5

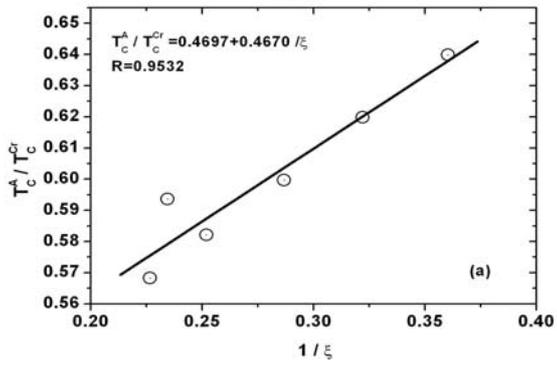

Fig.6a

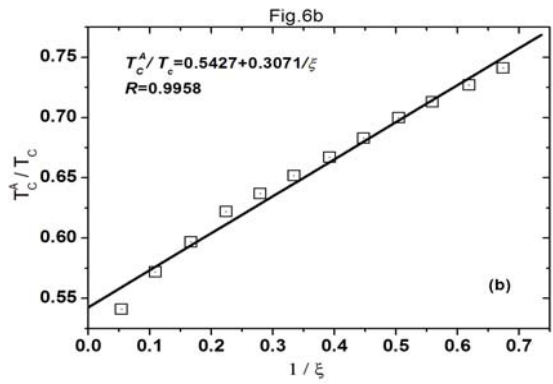

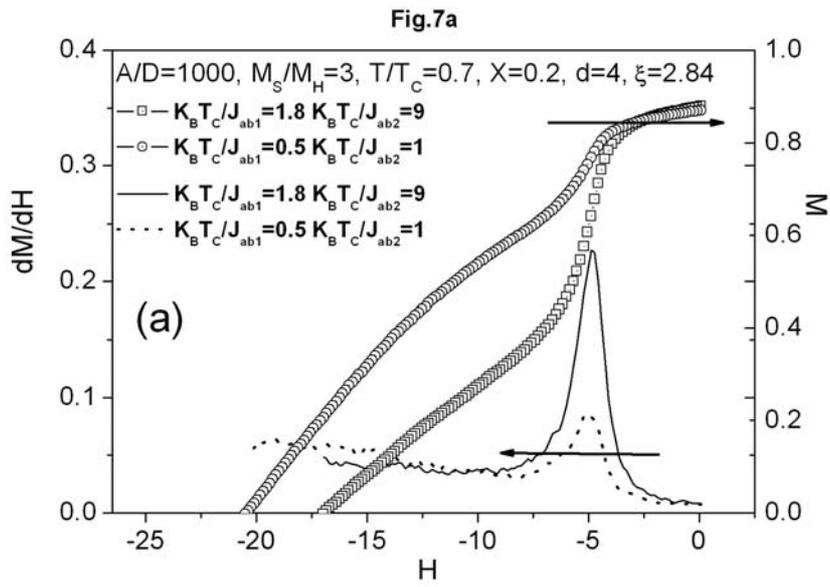

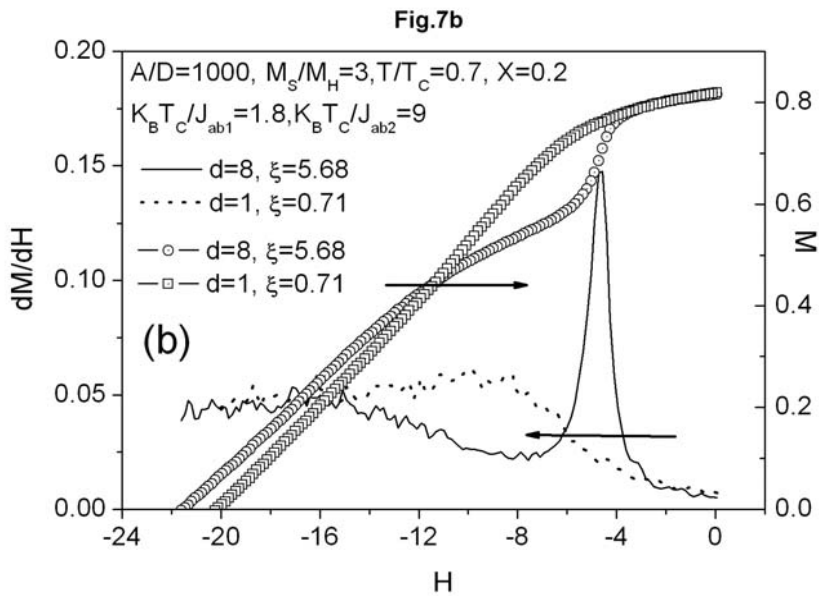

Table 1. Measured Curie temperature and other microstructure parameters in Finemet.

| d(nm) | Vc(%) | ξ(nm) | $T_c^A$ (°C) | $T_c^{cr}$ (°C) | $T_c^A/T_c^{cr}$ |
|---|---|---|---|---|---|
| 0 | 0 | ~ | 312.6 | ~ | ~ |
| 5 | 15 | 4.4104 | 345.3 | 607.6 | 0.5683 |
| 7 | 24 | 4.2640 | 355.0 | 598.0 | 0.5936 |
| 7.3 | 31 | 3.4862 | 357.3 | 595.8 | 0.5997 |
| 14.08 | 55 | 3.1050 | 371.2 | 598.8 | 0.6199 |
| 18 | 55 | 3.9694 | 348.3 | 598.3 | 05821 |
| 20.23 | 68 | 2.7752 | 372 | 581.3 | 0.6399 |

Table 2. Reported Curie temperature and other microstructure parameters in Fe-B-Nb-Cu in reference [8,9].

| Composition of alloy | d(nm) | ξ(nm) | $T_c^a$ (°C) | $T_c^{a*}$ (°C) | $l$(nm) | $T_c^{cr}$ (°C) | $T_c^a/T_c^{cr}$ |
|---|---|---|---|---|---|---|---|
| $Fe_{75.5}B_{19.2}Nb_{4.3}Cu_{1.0}$ | 9 | 12 | 273 | 220 | 0.58 | 768 | 0.36 |
| $Fe_{74.2}B_{20.2}Nb_{4.5}Cu_{1.1}$ | 10 | 11 | 298 | 225 | 0.73 | 624 | 0.48 |
| $Fe_{73.5}B_{20.7}Nb_{4.5}Cu_{1.2}$ | 10 | 7 | 343 | 235 | 0.44 | 1094 | 0.32 |
| $Fe_{70.7}B_{22.9}Nb_{5.1}Cu_{1.3}$ | 10 | 5 | 355 | 240 | 0.54 | 772 | 0.46 |
| $Fe_{69.5}B_{23.9}Nb_{5.3}Cu_{1.3}$ | 10 | 4 | 365 | 235 | 0.47 | 788 | 0.46 |
| $Fe_{69.5}B_{23.9}Nb_{5.3}Cu_{1.3}$ | 11 | 4 | 364 | 235 | 0.48 | 772 | 0.47 |
| $Fe_{67.1}B_{25.7}Nb_{5.7}Cu_{1.4}$ | 12 | 4 | 355 | 230 | 0.46 | 817 | 0.44 |